\documentclass[chi_draft]{sigchi}

 \toappear{
 }

\usepackage{amssymb,amsmath}
\usepackage{algorithmic}
\usepackage{algorithm}
\makeatletter
\renewcommand{\ALG@name}{Pseudocode}
\makeatother
\usepackage[tight,footnotesize]{subfigure}
\usepackage{url}

\usepackage{balance}       % to better equalize the last page
\usepackage{graphics}      % for EPS, load graphicx instead 
\usepackage[T1]{fontenc}   % for umlauts and other diaeresis
\usepackage{txfonts}
\usepackage{mathptmx}
\usepackage[pdflang={en-US},pdftex]{hyperref}
\usepackage{color}
\usepackage{booktabs}
\usepackage{textcomp}
\usepackage{caption}

\usepackage{microtype}        % Improved Tracking and Kerning
\usepackage{ccicons}          % Cite your images correctly!
\usepackage{todonotes}

\def\plaintitle{}

\def\emptyauthor{}
\def\plainkeywords{Authors' choice; of terms; separated; by
	semicolons; include commas, within terms only; required.}

\makeatletter
\def\url@leostyle{%
	\@ifundefined{selectfont}{
		\def\UrlFont{\sf}
	}{
		\def\UrlFont{\small\bf\ttfamily}
}}
\makeatother
\def\pprw{8.5in}
\def\pprh{11in}

\definecolor{linkColor}{RGB}{6,125,233}
\hypersetup{%
	pdftitle={\plaintitle},
	pdfauthor={\emptyauthor},
	pdfkeywords={\plainkeywords},
	pdfdisplaydoctitle=true, % For Accessibility
	bookmarksnumbered,
	pdfstartview={FitH},
	colorlinks,
	citecolor=black,
	filecolor=black,
	linkcolor=black,
	urlcolor=linkColor,
	breaklinks=true,
	hypertexnames=false
}

\usepackage{color}
\definecolor{highlight}{rgb}{1,1,0.6}
\definecolor{link}{rgb}{0.5,0.0,0.0}
\definecolor{cite}{rgb}{0.0,0.0,0.6}
\definecolor{url} {rgb}{0.3,0.0,0.3}
\definecolor{grey}{rgb}{0.3,0.3,0.3}

\usepackage{graphicx}

\usepackage[pdftex]{hyperref}
\hypersetup{%
	pdfauthor={blind submission}, %
	bookmarksnumbered, %
	pdfstartview={c}, %
	colorlinks,%
	citecolor=black, %
	filecolor=black, %
	linkcolor=black, %
	urlcolor=black}

\usepackage{soul}
\sethlcolor{highlight}

\begin{document}
\title{Mind My Value: a decentralized infrastructure for fair and trusted IoT data trading}

%\title{\plaintitle}

\numberofauthors{5}
\emptyauthor
\author{%
	\alignauthor{Paolo Missier\\
		\affaddr{Newcastle University}\\
		\affaddr{Newcastle, UK}\\
		\email{\textit{paolo.missier@newcastle.ac.uk}}}\\
	\alignauthor{Shaimaa Bajoudah\\
		\affaddr{Newcastle University}\\
		\affaddr{Newcastle, UK}\\
		\email{\textit{S.Bajoudah1@newcastle.ac.uk}}}\\
	\alignauthor{Angelo Capossele\\
		\affaddr{Digital Catapult}\\
		\affaddr{London, UK}\\
		\email{\textit{angelo.capossele@digicatapult.org.uk}}}\\
	\alignauthor{Andrea Gaglione\\
	\affaddr{Digital Catapult}\\
	\affaddr{London, UK}\\
	\email{\textit{andrea.gaglione@digicatapult.org.uk}}}\\
	\alignauthor{Michele Nati\\
	\affaddr{Digital Catapult}\\
	\affaddr{London, UK}\\
	\email{\textit{michele.nati@digicatapult.org.uk}}}\\	
}

\maketitle

\begin{abstract}
Internet of Things (IoT) data are increasingly viewed as a new form of massively distributed and large scale digital assets, which are continuously generated by millions of connected devices.
The real value of such assets can only be realized by allowing IoT data trading to occur on a marketplace that rewards every single producer and consumer, at a very granular level.
Crucially, we believe that such a marketplace should not be owned by anybody, and should instead fairly and transparently self-enforce a well defined set of governance rules.
In this paper we address some of the technical challenges involved in realizing such a marketplace.
We leverage emerging blockchain technologies to build a decentralized, trusted, transparent and open architecture for IoT traffic metering and contract compliance, on top of the largely adopted IoT brokered data infrastructure.
We discuss an Ethereum-based prototype implementation and experimentally evaluate the overhead cost associated with Smart Contract transactions, concluding that a viable business model can indeed be associated with our technical approach.
\end{abstract}

\section{Introduction}
Much of the expected value associated with the growing industry of Internet of Things (IoT) devices \cite{7004800} is to be found in the streams of data generated by those devices.
Application areas where interest in IoT data streams is growing range from health care \cite{7113786} to personal fitness, smart cities \cite{Perera2014}, optimization of energy consumption at home, and many more.
In each of these areas, the value of IoT is only delivered when the continuous data streams produced at the edge of the network are aggregated and analyzed by data consumer processes hereafter referred as Value Added Services (VAS).

Some of these applications are just emerging.
For example, in a public transport network like the London underground, the density of personal travel card swipes over time at individual metro stations may be
useful not only to the transportation authority, but also to taxi companies, which can benefit from the knowledge of any anomalous passenger traffic pattern, i.e., by placing their fleet at the right stations at the right time. A VAS that specializes in data analytics may therefore buy metro passenger data together with footfall data collected, for example, through an IoT infrastructure, and sell
recommendation services to taxi companies.
In response to this ``technology push'', new business models are indeed emerging \cite{Stahl2016,7765669} where data are viewed as tradeable digital assets. However, the lack of trust and  incentive in trading such assets is hindering their larger availability from producers to consumers.

In this paper we propose an initial technical infrastructure for a new kind of data marketplace that, in the long run, is designed to meet four main requirements.
First, the marketplace should be dynamic and flexible in order to enable the new and unanticipated kind of business relationships just illustrated. It should be possible to quickly  establish and then fulfil contracts between one and possibly many producers and the VAS, with guarantees of compliance and fairness.
Second, the marketplace should allow not only organizations, but also individuals to gain value from their data. 
For example, today it is possible to quantify an athlete's effort during a competition using a number of wearable devices, from bio-harness to accelerometers, to video feeds.
One can imagine that individuals may decide to let VAS access their data feeds, in return for some benefit (monetary or otherwise).
In the near future, athletes may be able to sell these feeds to followers who are interested in tracking their competition online.
There are examples in the UK today, where individuals get heavy discounts on smart watches from health insurance companies, provided they let the company access their fitness data.
Third, the main asset traded in the marketplace are streams of IoT data. This is not usual: a 2012 survey of data vendors \cite{Schomm2013}, for example, includes 46 data suppliers, however the definition of data marketplace used in the paper is generic (``a platform on which anybody can upload and maintain data sets, with license-regulated access to and use of the data'') and geared towards static data, like Microsoft's Azure Data Market.
In contrast, our requirement entails the typical ``Big Data'' challenges of high Volume, high Velocity, and high Variety of the streams.
Finally, it should be possible to run a completely decentralized marketplace which operates according to governance rules defining what kinds of contract and transactions are acceptable, and stipulating sanctions when the rules are violated. 
Contrary to existing proposals, e.g., \cite{Cao:2016:MMR:2926746.2883611}, we are going to assume  there is no central trusted authority appointed to enforce those rules. The assumption is that due to the unpredictable variety of actors trading  in such a marketplace, a multi-stakeholders decentralized trust will better adapt than a centralized one.
In this paper we focus specifically on this novel aspect. We investigate the use of blockchain (distributed ledger), and specifically of Smart Contracts \cite{Buterin2014}, as a technology enabler for an authority-free, trusted data trading infrastructure.

%In summary, we envision a decentralized, self-regulating IoT data marketplace where governance rules are automatically enforced and fairness of the transactions is regulated by means of a trust model amongst participants.

The contributions in the paper can be summarized as following.
\begin{itemize}
	\item We present a conceptual model for tracking brokered IoT data flows from gateways to VAS in the cloud, which embodies a methodology to achieve granular metering of IoT data trading.
	\item We explore the use of blockchain technology and Smart Contracts to remove the need for a centralized trust when settling contracts. 
	\item We present a proof-of-concept implementation of this trading infrastructure. We adapt the popular open source Mosquitto MQTT broker to add traffic metering capabilities, and use the Ethereum smart contracts technology for enforcing contract definition and trigger dispute resolution.
	\item We carry out an experimental evaluation identifying the viable boundaries for the prices of digital assets, which make the trading infrastructure economically sustainable.
	%relative to the number and frequency of transactions, taking account of the current Ethereum Smart Contract cost model. 
	We also assess the capability of Ethereum Smart Contracts to handle a stream of contract settlements at varying arrival rate, and conclude that they are indeed a viable option for the validation of contract compliance.
	\item As the use of Smart Contracts is a novel feature for any IoT architecture, we conclude the paper with a discussion on the challenges and lessons learnt from the use of this emerging and enabling technology.
\end{itemize}

\section{Brokered IoT data as tradeable assets}

We now present our conceptual model for the specification and enforcement of streamed data exchange agreements.
Following common IoT infrastructures, we are going to assume that data exchanges are mediated by one or more brokers.
Initially, we assume the brokers are trusted. In Sec. \ref{sec:no-trust} we are going to explore the consequences of relaxing this assumption.

\subsection{Contracts and pricing}

Let  $P = \{p_1 \dots p_n \}$ and $C = \{ c_1 \dots c_m \}$ denote the set of producers (IoT devices) and consumers (VAS) that participate in the trading, respectively.
In the standard publish/subscribe model for data brokering, the $p_i$ act as publishers and the $c_j$ are subscribers.
These participants agree on a set $T = \{ t_1 \dots t_r \}$ of topics.
In IoT data brokering, messages are generated by gateways, which are responsible for segmenting raw data streams from edge devices into discrete messages.
The topic associated with each message describes the type of data stream, for example ``heart rate'', ``GPS track'', ``glucose reading'', ``energy reading'', etc.
Suppose $ p_i $ publishes data on a set of topics $T_{i} \subset T$.
A consumer $ c_j  $ enters into a \textit{contractual agreements} with a producer $ p_i  $ by subscribing to a subset $T_{ij} \subset T_i$ of the topics available from $p_i$, possibly only for the duration of a time window $ W = [w_s, w_e] $.
Such an agreement is interpreted as ``$p_i$ agrees to let $c_j$ receive a copy of all its messages tagged with any $t \in T_{ij}$ during $W$, and $c_j$ agrees to pay a corresponding data exchange fee. 
The broker manages all subscriptions and is responsible for reliably delivering to $ c_j  $ a copy of each message that has a topic that $ c_j $ subscribes to.
Note that in the standard pub/sub model, publishers and subscribers are unaware of one another, and their interaction is entirely mediated by the broker. 
However, it is easy to extend the model by assuming that the broker will only deliver messages from $ p_i $ to $ c_j $ if $ c_j $ has an active agreement (i.e., relative to $W$) with $ p_i $.\footnote{This can be easily realized in a MQTT-based broker, which we use in our implementation, e.g., by encrypting payload data.}

%for instance by embedding a device name into a topic tree
%	
%	. When topics are into a hierarchy, 
%
%\textcolor{blue}{To enable explicit marketplace} interactions, we propose a variation of the model whereby every message topic embeds the publisher's ID, which in combination with MQTT's topic regular expressions capability enables a subscriber to selectively choose the stream they intend to license.
%For example, device ``1234'' may publish messages about a specific room's temperature using topic  \#/temperature/1234/living\_room\_temperature. 
%A subscriber may choose to subscribe to this specific topic, or to any  living\_room\_temperature message using expression \#/temperature/*/living\_room\_temperature, or to any generic temperature message from ``1234'' or from any publisher.

%Note incidentally that the model includes the case, not shown in the figure, where a VAS may aggregate data from multiple $ p_i $s and then publish this value-added data, enabling other VAS to license it. 
%In this case, a VAS is both a consumer and a producer.

%While a variety of pricing models have recently been proposed for digital assets in emerging data marketplace scenarios \note{CITE}, discussing how the marketplace sets the data prices is beyond the scope of this work. Instead, we are simply going to assume that each individual message is a digital asset with a constant unit value $\mathit{val}(t_k)$, which is determined solely by the message's topic $t_k$.
A variety of pricing models have recently been proposed for digital assets in emerging data marketplace scenario \cite{Sen:2015:SDP:2847579.2756543,Li:2014:TPP:2691190.2691191,7553037,7437020}.
In this work we are going to assume a simple model where each individual message has a constant unit value $\mathit{val}(t_k)$, which is determined solely by the message's topic $t_k$. 
While our infrastructure is largely agnostic to the specific data pricing model, in our evaluation we analyze the economic sustainability of a decentralized marketplace. Specifically, in Sec.\ref{sec:evaluation} we analyze the cost overhead of enforcing agreements given the current cost model associated with Smart Contract transactions.

\subsection{Data traffic cubes and centralized settlement}

Contract enforcement and settlement involves calculating the total price associated with the messages that have been routed from each $ p_i $ to each $ c_j $ within each $W$.
Since we have assumed that the price is  determined only by the number of messages and the unit cost for each topic, this simply requires keeping a count of the number of messages about topic $t_k$ that originated from $p_i$ and reached $c_j$ during $W$, grouped by $ p_i $, $ c_j $, and $ t_k $.
We denote each of these counts as $N_{ijk}(W)$.
Generating these counts requires the broker to be capable of \textit{metering} all traffic, that is, of logging all messages.
The log consists of a set of tuples:
$ \{\langle p_i, c_j, t_k \rangle  \} $
At the end of each $W$, the log is aggregated over each $p_i \in P, c_j \in C, t_k \in T$, resulting in a set of tuples that we call a \textit{traffic cube}:
\begin{equation}\label{eq:cube}
\mathit{cube}(W) = \{ \langle p_i, c_j, t_k, N_{ijk}(W) \rangle \}_{p_i \in P, c_j \in C, t_k \in T}
\end{equation}

Here we borrow our terminology from standard database practice (OLAP, or Online Analytical data Processing), where a  ``cube'' is a table with $ N $ attributes, in which the first $ N-1 $ attributes are  dimensions in a database schema (in our case, these are the Producers, Consumers (the VAS), and Topics) and the last is an aggregation over values in the database for each combination of the dimensions--in our case, a count.
We use a matrix indexing notation to refer to specific cells in the cube, i.e.:
\[ \mathit{cube}(W)[p_i, c_j, t_k] = N_{ijk}(W) \] 
These cubes contain summaries of all data flows observed by a broker. Notice that they only contain \textit{metadata}, i.e., the counts, but not the content of the messages.
Note that the values in the cube may be sparse, i.e., $N_{ijk}(W) = 0$ whenever $c_j$ does not subscribe to $t_k$.

\textit{Settlement} is the process of calculating the total fee owed by each $ c_j $ to each $ p_i $ at the end of each $W$.
This is computed by suitably aggregating the counts in the cube, namely:
\begin{equation}
\mathit{fee}(c_j, p_i, W) = \sum_{t_k \in T} N_{ijk}(W) \cdot \mathit{val}(t_k)
\label{eq:balance}
\end{equation}
and the total profit for $ p_i  $ during $W$ is
\begin{equation}
\mathit{profit}(p_i, W) = \sum_{c_j} \mathit{fee}(c_j, p_i, W)
\label{eq:reward}
\end{equation}
In the centralized scenario we have considered so far, settlement is straightforward, as the broker is entrusted with generating accurate logging and thus complete and correct cubes.
Note that, under the same trust assumptions, settlement extends easily to a more realistic scenario where multiple brokers are deployed, each enhanced with the same logging capabilities and local traffic reporting service.
However, settlement becomes challenging in an extended model where there is no assumption of trust in the broker.
In this case, fee settlement must rely on data traffic counts that are calculated independently by each participant, based on the portions of data flows that are visible to each of them, with the 
further complication that participants cannot be trusted to generate complete and correct cubes.
This decentralized scenario is illustrated in Fig. \ref{fig:iot-tracking-arch-2} and discussed in the next Section.

\section{Removing the need  for trust in the network}  \label{sec:no-trust}

\begin{figure*}[!ht]
	\centering
	\captionsetup{justification=centering, margin=2cm}
	\includegraphics[width=0.52\textwidth]{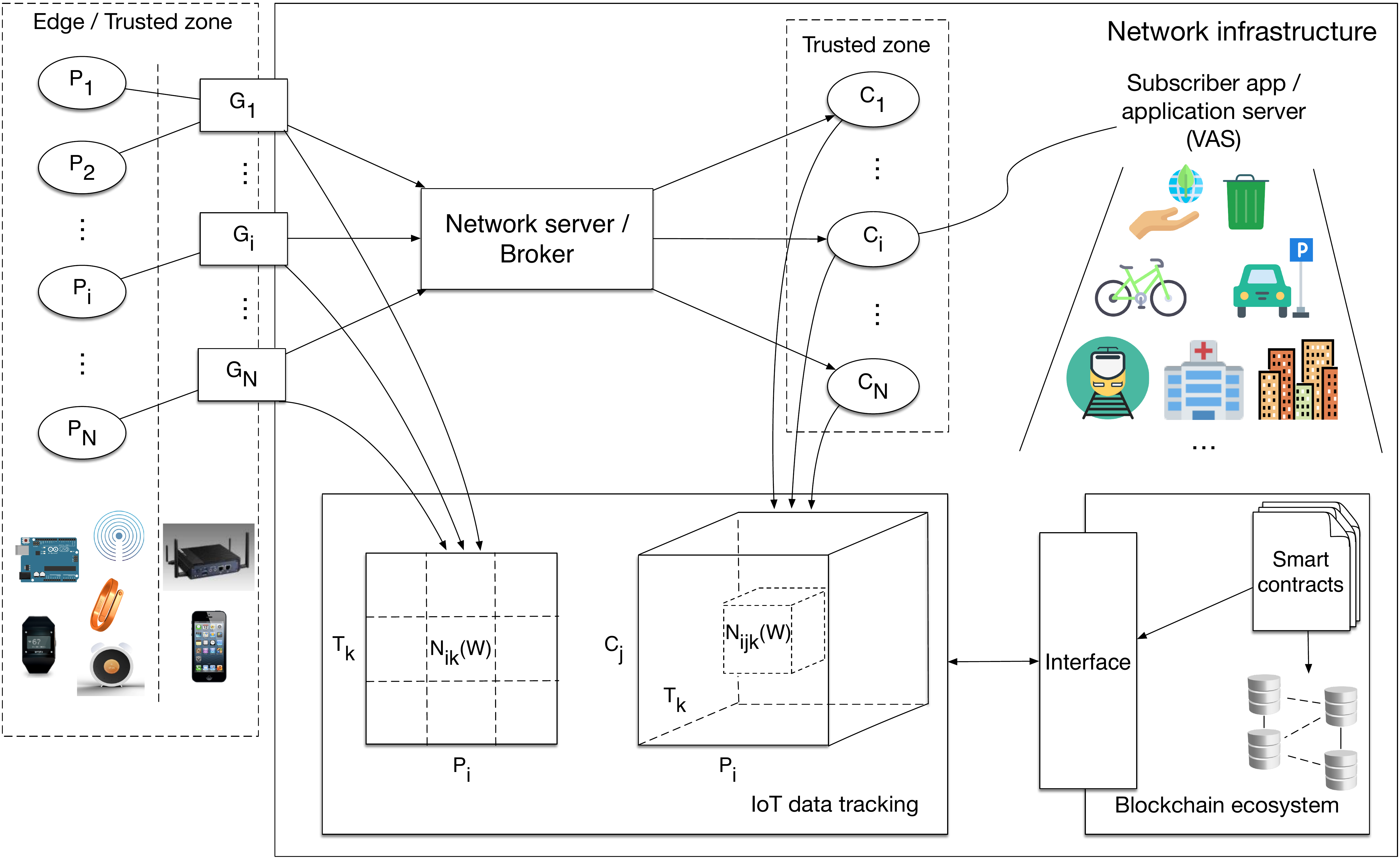}
	\caption{Blockchain and Smart Contracts based architecture for decentralized metering of IoT data trading between Producers (P) and Consumers (C).}
	\label{fig:iot-tracking-arch-2}
\end{figure*}

A trading where the reward model is based on message counts is vulnerable to malicious behavior.
Specifically, producers have an incentive to claim to have produced more messages than they have in reality, while conversely, consumers (the VAS) have an incentive to under-report the number of messages they receive.
When we remove the assumption that the brokers are trusted, we must also accept that the brokers may collude with any of the participants, and thus deliver traffic cubes that may not be correct or complete.
Discovering such collusions may not be possible when the broker is the only source of traffic counts available to the settlement service.  At the same time, resolving any disputes amongst pairs of participants requires a public and irrefutable record of the reported traffic.
To address these problems, we rely on two overarching principles: (1) personal responsibility of each participant in the trading, which shall report their own counts of messages sent (publishers) or received (subscribers) using \textit{trusted zones} (see Fig.\ref{fig:iot-tracking-arch-2} and description below),
 and (2) transparency, whereby these reports are posted as part of immutable and verifiable blockchain transactions.
These principles translate into a two-steps approach.

Firstly, we remove the assumption that traffic cubes are generated by the broker alone, and instead enable networks elements close to the publishers and to the subscribers, i.e., gateways and VAS respectively, to generate the cubes. This is shown in Fig. \ref{fig:iot-tracking-arch-2}.
Secondly, we adopt emerging consensus-based distributed transaction ledgers, specifically blockchain and Smart Contract technologies, to realize the settlement service.
As we explain in more detail later (Sec. \ref{sec:blockchain}), Smart Contracts extend the standard blockchain transaction model by adding the capability to execute arbitrary code, which operates on data structures contained in the transaction itself. 
In this case, a blockchain transaction that is initiated at the end of each window $W$ may operate on the collection of traffic cubes that participants make available at the end of $W$.
This  approach provides at the same time transparency and accountability, because the content of the blockchain is public and can be inspected, and a way to address disputes, because for each W, multiple (partial) views of each cube are made available to the settlement service.

\subsection{Unilateral traffic cubes} \label{sec:u-cubes}
%\ag{This sub-section now includes the architecture. It should be integrated in the previous main Section}
Traffic cubes that are generated by the broker summarize the entire traffic during $W$.
In contrast, traffic summaries generated by trading participants reflect the \textit{local} views of each participant in the data exchanges.
These are therefore necessarily partial and incomplete, as each participant, unlike the broker, has no visibility of the end-to-end data flows. 
We denote these as \textit{unilateral} traffic cubes, defined as follows.
Let us assume that a producer does not know which VAS subscribe to its stream, while subscribers know the source of the messages they receive. 

Let $\mathit{sub}(t_k) \subseteq C $ denote the set of subscribers to $t_k$.
A \textit{publisher's cube} $\mathit{cube}^p$ is a slice of a complete traffic cube, for a specific producer $p_i$ and without the consumer dimension:
\[
\mathit{cube}^p(W, p_i)  =  \{ \langle t_k,  N^s_{ik}(W) \rangle \}_{t_k \in T}
\]
where $N^s_{ik}(W)$ is the count of messages with topic $t_k$ sent by $p_i$ during $W$.
Note that $ p_i $ can compute $ N^s_{ik}(W)  $ from its own data flow log, but not $ N_{ijk}(W)  $.

As subscribers know the source of the messages they receive, we may assume that a subscriber will produce summary reports that include the publisher dimension, but which only contain the tuples that pertain to a single $c_j$. Thus, a \textit{subscriber's cube}, $ \mathit{cube^s} $ is defined as:
\[
\mathit{cube^s}(W, c_j)  =  \{ \langle p_i, t_k, N_{ijk}(W) \rangle | c_j \in \mathit{sub}(t_k)\}_{p_i \in P, t_k \in T}
\]

Figure 1 concretely illustrates this setting. To remove the need for a centralized trust, we push it towards the borders of the data flow network by defining two \textit{trusted zones}.
The first trusted zone includes all the elements at the edge of the network infrastructure, such as the IoT devices $P$ and the gateways $ G_i $, whereas the second one includes $ C $.
%The latter range from monitoring, to health applications, assets tracking, smart city applications, etc.
IoT data are still routed towards the VASs through brokers--using a publish-subscribe pattern--or network servers.
However, we now assume that a new, independent IoT data tracking component receives the unilateral cubes by gateways and VASs.
%Finally, a Smart Contract deployed on a blockchain periodically accesses the traffic cubes to realize settlement services and resolve possible conflicts.
Finally, a Smart Contract, decentralized trusted service deployed on a blockchain, periodically accesses the traffic cubes to realize settlement services and resolve possible conflicts.

\subsection{Consistency and settlement with unilateral cubes}

Suppose that, at the end of $W$, every $p_i$ and $c_j$ produce  unilateral cubes relative to $W$. 
These form the set
\begin{equation}\label{eq:all-cubes}
\{ \mathit{cube}^p(W, p_i) \}_{p_i \in P}\cup \{\mathit{cube^s}(W, c_j) \}_{c_j \in C} 
\end{equation}
As each of these cubes provides a partial view of the same complete cube $ \mathit{cube}(W) $ that would have been generated centrally by a broker, we expect that the values found in these cubes be somehow consistent with $ \mathit{cube}(W) $.
The pub/sub model implies that the number of messages sent by $p_i$ with topic $t_k$ during $W$ must be equal (assuming no messages are lost and ignoring duplicate transmissions, as in MQTT QoS level 3) to the number of messages each $c_i$ that subscribes to $t_k$ receives from $p_i$. 
We can capture this constraint formally using our cubes notation, as follows.
For each $ p_i \in P, t_k \in T, c_j \in \mathit{sub}(t_k)$:
{\begin{equation}\label{eq:cubes-consistency}
\begin{split}
& \mathit{cube}^p(W, p_i)[t_k] = N^s_{ik}(W) = \\
& \mathit{cube}(W)[p_i, t_k, c_j]  = N_{ijk}(W) = \\
& \mathit{cube^s}(W, c_j)[p_i, t_k]
\end{split}
\end{equation}
%\vspace{-10pt}
%\begin{definition}
We say that the set (\ref{eq:all-cubes}) of all unilateral cubes is \textit{consistent} at $W$, if and only their contents satisfy constraint (\ref{eq:cubes-consistency}).
%\begin{end}
We use this definition as a basis for settlement of message exchanges within each $W$, in the general case that the broker cannot be trusted to provide a single global cube that is complete and correct.
Specifically, in our architecture we  now assume that a new, independent component receives all cubes in (\ref{eq:all-cubes}) at the end of each $W$, and checks their consistency using (\ref{eq:cubes-consistency}). 
In the next section we discuss a practical implementation of this idea, where this new component is realized as an Ethereum Smart Contract and unilateral cubes are posted publicly as part of blockchain transactions.
In this decentralized scenario, such a settlement service must be able to deal with two inter-dependent issues, namely (a) \textit{completeness} and (b) \textit{consistency} of the set (\ref{eq:all-cubes}) of all cubes.
The case when set (\ref{eq:all-cubes}) is both complete and consistent is straightforward and results in successful settlement, as all information for settlement is available, and there are no disagreements.

When the set of cubes is incomplete, we may try to use (\ref{eq:cubes-consistency}) to propagate missing values from the more complete to the less complete cubes. 
More precisely, suppose $ \mathit{cube}^p(W, p_i) $ is missing for a $p_i$.
If $ N_{ijk}(W) = \mathit{cube^s}(W, c_j)[p_i, t_k] $ is available for some $t_k$ and some $ c_j \in \mathit{sub}(t_k) $, then we set $  \mathit{cube}^p(W, p_i)[t_k]=  N_{ijk}(W) $.
In practice, this can be viewed as ``taking $ c_j $s word for $p_i$s missing report''.

Symmetrically, the settlement service may use the available $ \mathit{cube}^p(W, p_i)$, in combination with subscription information $ \{\mathit{sub}(t_k) | t_k \in T \}$, to fill in missing values in 
$  \mathit{cube^s}(W, c_j)  $, i.e., by setting 
$ \mathit{cube^s}(W, c_j)[p_i, t_k]  =  \mathit{cube}^p(W, p_i)[t_k]$ for each $t_k$ and each $c_j \in \mathit{sub}(t_k)$.
Of course, there is no guarantee that all missing values can be propagated. 
In this case, settlement for the $\langle p_i, c_j \rangle$ pairs corresponding to the missing cube entries is simply not possible.

The final, and perhaps most important case occurs when constraint (\ref{eq:cubes-consistency}) is violated for some combination of $\langle p_i, c_j, t_k \rangle$.
This may be due to the malicious cases of over-reporting producers, or under-reporting subscribers.
Either of these  scenarios manifests itself as inequalities in (\ref{eq:cubes-consistency}), of the form:
\begin{equation}\label{eq:inconsistencies}
\mathit{cube}^p(W, p_i)[t_k] > \mathit{cube^s}(W, c_j)[p_i, t_k]
\end{equation}
In this situation, we are able to detect the inconsistency, but we may not have enough information to determine whether $p_i$, $c_j$, or both are guilty of fraud.
Such determination is beyond the scope of this paper, but in the final discussion section we present initial ideas on promoting a self-regulating exchange infrastructure in the presence of such unresolvable inconsistencies.
In our initial implementation, described next, the settlement service simply reports the detected inequalities.

\section{Initial proof-of-concept realization}  \label{eq:realisation}

%\note{PM: below are the spare bits of implementation that I removed from the first part}
%
%\authornote{During normal operation, the broker In our testbed we by extending a Mosquitto MQTT broker so it logs each message routing event as a tuple into a \textit{TrackerDB} database
%	$ \langle p_i, c_j, t_k \rangle $
%}
%
%\authornote{We use a Cassandra NoSQL database for the \textit{TrackerDB}, to ensure scalability. 
%	A traffic reporting service generates traffic cubes on demand by querying the \textit{TrackerDB}, in response to requests issued by client applications (including possibly independent third party clients). The service is accessible through a REST interface.}
%
%\authornote{In our testbed deployment we use MQTT brokers, and the gateways generate messages by encapsulating either a batch of the raw data stream or an aggregation of it, depending on the type of stream, into the MQTT payload.}

\subsection{Background concepts: Blockchain and Smart Contracts}  \label{sec:blockchain}
Blockchain is essentially a distributed ledger of information (e.g., a transaction from A to B in the bitcoin world), a copy of which cannot be arbitrarily altered without being spotted and for which consistency of each information can be achieved through a decentralized and distributed consensus, without requiring trust in any third party but instead, through large and flat pool of so-called miners using cryptographic primitives~\cite{nakamoto2008bitcoin}.
Blockchain has been later leveraged to manage Smart Contracts, small pieces of software that encode a set of conditions and actions that a machine can interpret and that can be executed as expected using the blockchain infrastructure without third party involvement or supervision~\cite{Buterin2014}. Smart Contracts represent therefore a well-suited decentralized tool to implement cube consistency and settlement functionalities. Being one of the most adopted and well-supported by the developers community, we decide to use Ethereum Smart Contract implementation\footnote{https://www.ethereum.org}.

In the Ethereum network, any node uses a virtual machine (EVM), which can run code of arbitrary algorithmic complexity, to execute smart contracts, the integrity of whose is always guaranteed. A smart contract can perform various state updates and account balancing.  
Executing a smart contract results in one or more transactions to be validated. Each transaction has a cost (e.g., fee) associated, which translates into incentive for any miner within the network to independently execute it. 
More specifically, any operation being performed within a transaction consumes a fixed amount of Gas. Miners fees are therefore proportional to the amount of Gas used. Gas price is measured in terms of Ether (the Ethereum cryptocurrency). Every transaction specifies the Gas price a smart contract is willing to pay for its execution, thus, the total fees paid for a transaction is the result of Gas amount multiplied by the Gas selected price.

\vspace{-10pt}
\subsection{Implementation}
\label{sec:implementation}
For the purpose of experimentation and evaluation, we have adapted the open-source Mosquitto MQTT broker to support message logging and cubes generation into a Cassandra NoSQL database. We refer to it as the \textit{TrackerDB}. We connected to the MQTT broker real producers using channels provided by the ThingSpeak platform\footnote{https://thingspeak.com}.
Using the TrackerDB, we are able to simulate the generation of unilateral cubes that can be either complete and correct, or reflect malicious behaviour, for evaluation purposes.
The TrackerDB can be queried by any third party client through a REST service interface. Smart Contracts interact with the service through an Ethereum-specific mechanism, described below. 
In reality, unilateral cubes would be generated by gateways on the producers side as well as by VAS within their trusted zones. This does not affect the properties of the cubes compliance and settlement, because liability is pushed at the edge.
%
%Section~\ref{sec:no-trust} provides some detail on how unilateral cube generation can be implemented. 
%by extending a Mosquitto MQTT broker so it logs each message routing event as a tuple into a \textit{TrackerDB} database $ \langle p_i, c_j, t_k \rangle $

We now focus on the use of Smart Contracts in this setting. We developed them using Solidity, the Ethereum's scripting language. To implement the contracts, we assigned an Ethereum account to each producer and VAS. We connected these accounts to our private Ethereum test network, deployed on a single node with 6-core Intel Xeon E5-2640 and 16GB of RAM. We wrote, deployed and evaluated Smart Contracts in the network by using the Ethereum web browser based IDE Remix, connected to our private chain through Remote Procedure Call (RPC) protocol. In our implementation, accounts prepare and send the transactions to the blockchain to instances of Geth\footnote{https://github.com/ethereum/go-ethereum/wiki } through RPC. To measure Gas consumption, we used the debug tool provided by Remix and we observed the difference in the account balance before and after invoking a settlement contract.
A limitation of Ethereum smart contracts is that they cannot directly access off-chain data about real-world state and events. In our case this represents a challenge in acquiring unilateral cubes value. More precisely, Smart Contracts are independently executed by any node in the chain, thus, each execution needs to retrieve such information from an off-chain source independently, without any assurance on the information integrity.
To overcome this limit, the concept of \emph{oracle} has been introduced. Simply speaking, an oracle is a special contract that serves data requests from traditional contracts, by sourcing them from designated data feeds. 
Two options are possible for implementing oracles. The first one is relying on existing proxy services. Oraclize\footnote{http://www.oraclize.it} provides a \emph{programmable} oracle that can interact with any data source selected among a predefined set of standard channels. In addition Oraclize provides an authenticity proof by means of a TLSNotary proof which guarantees the authenticity and integrity of the retrieved data. These functionalities come at a cost. For each off-chain query, Oraclize requires a fee which includes a commission, ranging from 0.01\$ to 0.04\$, and a refund of the Gas used to perform the transaction. 
The other option is when each party of the contract, producer and VAS, independently update their view of unilateral cubes by pulling their values from cubes generator located within their trusted zones and then creating a transaction which embeds the cubes in the blockchain. This way any node executing the smart contract will have the same copy of that cube. As a result, costs associated to the use of an external oracle proxy, such as Oraclize, can be saved. Since in our model the responsibility and liability of producing faulty cubes is placed to producers and VAS, this option well suffices our needs.

Pseudocode~\ref{settlement_contract} shows the pseudocode of our settlement contract. For the sake of simplicity, this code snippet only accounts for the single producer and the single VAS scenario, although generalization is straightforward.
The contract first requests the involved parties to provide their unilateral cubes; then it uses this information to perform the actual settlement, by combining the two unilateral cubes. If the processed combined cube is consistent then a payment to the producer is performed, otherwise, a dispute resolution mechanism should be invoked\footnote{At this time, our implementation simply reports and log the detected inequalities.}.
\begin{algorithm}
	\caption{Cube settlement contract}\label{settlement_contract}
	\begin{algorithmic}
		\IF{sender $\neq$ authorizedAddress}
		\STATE throw
		\IF{queryId $=$ producerQuery}
		\STATE producer $\gets$ unilateralCube
		\STATE vasQuery $\gets$ update()
		\ELSIF{queryId $=$ vasQuery}
		\STATE vas $\gets$ unilateralCube
		\IF {producer $=$ vas}
		\STATE transfer(producerAccount, dataPrice, cube)
		\ELSE
		\STATE disputeResolution()
		\ENDIF
		\ENDIF
		\ENDIF
	\end{algorithmic}
\end{algorithm}
When a dispute resolution is invoked, payments are retained from being performed due to impossibility to clearly identify the correct unilateral cube. 
A reputation mechanism can be implemented in order to penalize both parties involved in a given settlement transaction and to promote them when a honest behavior is identified. 
As it is not expected that reputation computation will require off-chain interactions~\cite{schaub2016trustless,carboni2015feedback}, we are confident that not considering its implementation in this phase will not significantly affect the overall contract execution cost.

Table~\ref{tab:execution_costl} shows the execution cost of cube settlement operations expressed in Gas without and with Oraclize respectively. The most expensive operation to be performed is the \emph{contract deployment}, consuming from 175000 Gas without Oraclize to 2061490 Gas with Oraclize. The difference between these values is due to the higher number of functionalities implemented within Oraclize's API that the contract has to deploy\footnote{It is worth noticing that most of such functionalities are not required in a distributed liability model as the one promoted in our architecture}. Both the \emph{update} and \emph{callback} operations have a higher cost due to the Oraclize's fee, whereas the \emph{transfer} operation has the same cost.

\begin{table}[]
	\centering
	\caption{Execution cost of cube settlment contract operations.}
	\label{tab:execution_costl}
	\tiny 
	\resizebox{.4\textwidth}{!}{%
		\begin{tabular}{l|l|l|}
			\cline{2-3}
			& \multicolumn{2}{c|}{\textbf{Gas used}} \\ \hline
			\multicolumn{1}{|l|}{\textbf{Operation}} & w/o Oraclize & w Oraclize \\ \hline
			\multicolumn{1}{|l|}{Contract deployment} & 175000 & 2061490 \\ \hline
			\multicolumn{1}{|l|}{Update} & 41000 & 120000 \\ \hline
			\multicolumn{1}{|l|}{Callback} & 23000 & 70000 \\ \hline
			\multicolumn{1}{|l|}{Transfer} & 21000 & 21000 \\ \hline
		\end{tabular}%
	}
\end{table}
\vspace{-10pt}

\section{Evaluation and Lessons learnt}
\label{sec:evaluation}
Aim of this section is to quantify the cost of the smart contract described above and the associated cube settlement operations. By considering the scenario in which one VAS consumes the data of one producer, we evaluate how the cost of performing such contract affects the data price when the number of data exchanged and settlements transactions required changes. Considering different quantity of exchanged data reflect the different purpose of the exchange (event-based data rather than real-time series acquisition). Nevertheless, the reason for considering a variation in the number of required settlements needs some clarification. The most natural strategy will be to perform the settlement at the end of each contractually agreed data exchange, however in the early stage of an hypothetical marketplace where new producers and VAS join without necessarily trusting each other nor having an already established reputation, two situations might occur:
\begin{itemize}
	\item Producers and VAS have low reputation, hence, their trust level is low and the risk of claiming wrong unilateral cubes is high. By performing more than one cube settlement, in an initial rump-up phase of a given data exchange, will allow them to mutually increase their reputation and trust;
	\vspace{-5pt}
	\item Producers and VAS have high reputation, hence, they are expected to act honestly. Cube settlements may occur at a lower rate, only at the end of a data exchange phase, because the risk of producing faulty cubes is mitigated.
\end{itemize}
%1- 
%2- 

By evaluating the cost of performing the settlement operations, we are able to define the minimum price that VAS should pay for each consumed data in order to sustain the settlement infrastructure and eventually generate profit for the producers.
We define the minimum data price as the amount of Ether needed to at least cover the cost of contract deployment and transactions for performing cubes settlement operations. This means that if a producer sells data at the minimum price, its profit will be zero. 
At the time of writing, one Ether costs 220\$, however, its price is still very volatile.\footnote{\url{http://etherscan.io/chart/etherprice}}% (\url{etherscan.io/chart/etherprice}).
As result, transaction cost may frequently vary, thus leading to uncertainty about the economic feasibility of a specific application. We analyzed the capability of Ethereum to support a stable transaction cost by tuning the Gas price. The main drawback when setting a low Gas price is the increase of time required before a transaction is validated. Assuming a range of Gas price between 0.9 Gwei and 20 Gwei (9e-10 and 2e-8 Ether respectively), as minimum and average reported by the Ethereum network in 2017, the time required for a transaction to be validated in the chain varies from 2 minutes to 14 seconds (\url{etherscan.io/chart}). As explained before, even in the case of multiple settlements, we do not expect that meaningful data exchange will last less than 2 minutes, thus we consider a viable choice to select the current minimum Gas price.

Figure~\ref{fig:overall} shows a general overview of the minimum data price by varying the frequency of cubes clearance operations and the amount of transferred data. The price is directly proportional to the number of cubes settlement performed while inversely proportional to the data amount exchanged. Clearly, the more the data a VAS purchases, the less impacts the cost of performing cubes settlement. Depending on the type of data exchanged and trustworthiness of involved parties, this figure clearly shows how an optimal settlement strategy can always be found to dynamically adapt the number of settlement operations.

\begin{figure*}[t]
	\centering
	\subfigure[Minimum data price]{
		\includegraphics[width=.33\textwidth]{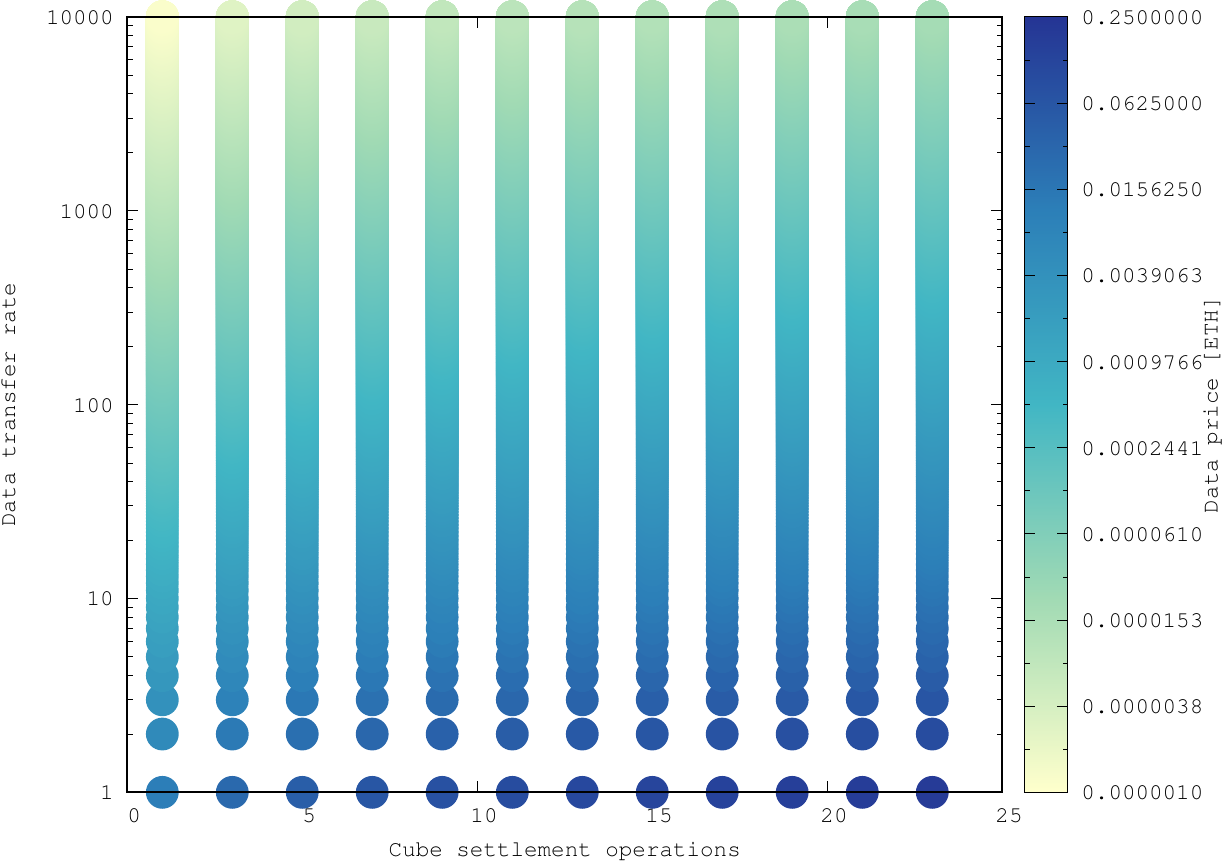}
		\label{fig:overall}
	}%
	\subfigure[Cost without Oraclize]{
		\includegraphics[width=.33\textwidth]{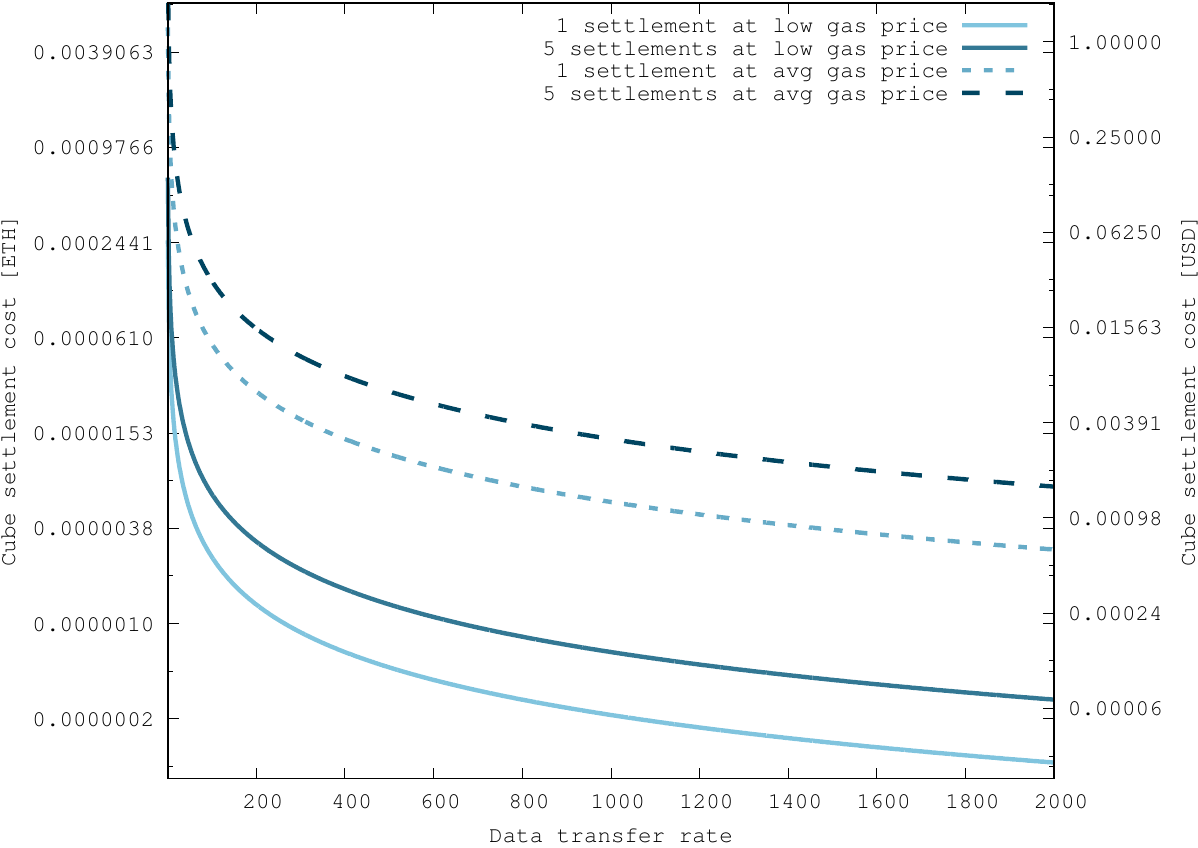}
		\label{fig:noOraclize}
	}%
	\subfigure[Cost with Oraclize]{
		\includegraphics[width=.33\textwidth]{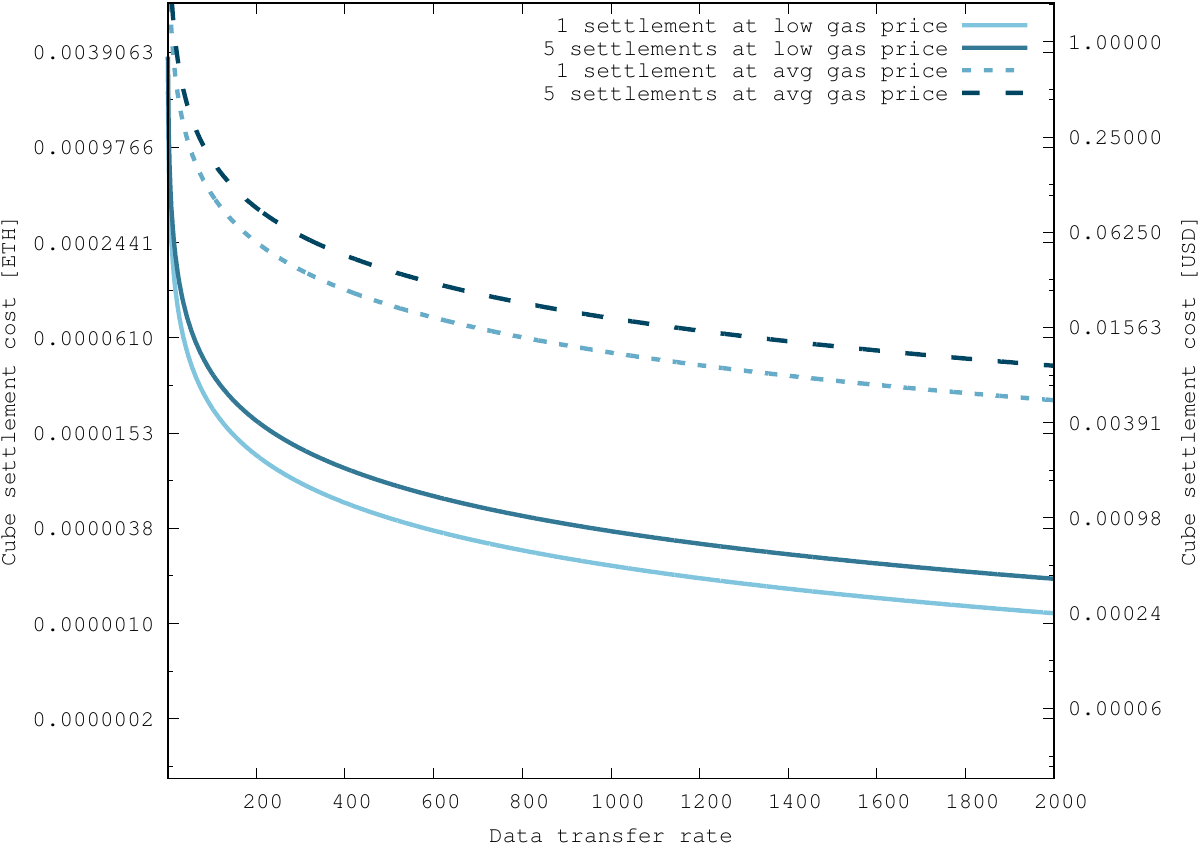}
		\label{fig:withOraclize}
	}%
	\caption{Cost of performing cube settlement operations for different data transfer rate.}
	\label{fig:all}
\end{figure*}

Figures~\ref{fig:noOraclize} and~\ref{fig:withOraclize} show the total cost of performing 1 or 5 cubes settlement operations for a fixed amount of transferred data, when considering a Gas price ranging from 0.9 Gwei to 20 Gwei.
More specifically, figure~\ref{fig:noOraclize} shows the case when each party of the contract use its oracle implementation, while figure~\ref{fig:withOraclize} shows the case when Oraclize functionalities are used. 
It is worth noticing the large costs increase (on average 4 times more), due to the commission and refund of the Gas used to perform the transaction to be paid to Oraclize. 
Without Oraclize, the cost of a single cube settlement transaction ranges from 9.9e-5 (\$ 2.18e-2) Ether to 2.2e-3 Ether (\$ 4.84e-1) when the Gas price selected is 0.9 Gwei and 20 Gwei, respectively. The more amount of data is transferred, the less impact the transaction cost has per single data. In fact, when performing a cube settlement operation spread over 2000 data, its cost ranges from 1.26e-7 Ether (\$ 2.77e-5) to 2.8e-6 Ether (\$ 6.16e-4). Alternatively, when performing 5 cube settlement over 2000 data, their cost ranges from 3.15e-7 Ether (\$ 6.93e-5) to 7e-6 Ether (\$ 1.54e-3).
%1.26e-07	3.15e-07	2.8e-06	7e-06
When using Oraclize, the cost of a single cube settlement transaction ranges from 3.61e-4 Ether (\$ 7.94e-2) to 8.02e-3 Ether (\$ 1.76) when the Gas price selected is 0.9 Gwei and 20 Gwei, respectively. When performing a cube settlement operation spread over 2000 data, its cost ranges from 1.11e-6 Ether (\$ 2.44e-4) to 2.46e-5 Ether (\$ 5.42e-3). Alternatively, when performing 5 cube settlement over the same amount of data, their cost ranges from 1.83e-6 Ether (\$ 4.03e-4) to 4.07e-5 Ether (\$ 8.95e-3).
%\note{Add oraclize graph description}
%1.1081205e-06	1.8299205e-06	2.46249e-05	4.06649e-05
\begin{table}[!htp]
	\centering
	\caption{Estimated data price for different use cases.}
	\resizebox{.4\textwidth}{!}{%
		\begin{tabular}{r|c|r|r|r|}
			\cline{2-5}
			\multicolumn{1}{l|}{} & \multicolumn{4}{c|}{\textbf{Data price}} \\ \cline{2-5} 
			\multicolumn{1}{l|}{} & \multicolumn{2}{c|}{w/o Oraclize} & \multicolumn{2}{c|}{w Oraclize} \\ \hline
			\multicolumn{1}{|c|}{\textbf{Data rate}} & ETH & \multicolumn{1}{c|}{USD} & \multicolumn{1}{c|}{ETH} & \multicolumn{1}{c|}{USD} \\ \hline
			\multicolumn{1}{|r|}{high} & \multicolumn{1}{r|}{5.73e-8} & 1.26e-5 & 2.09e-7 & 4.59e-5 \\ \hline
			\multicolumn{1}{|r|}{medium} & \multicolumn{1}{r|}{3.44e-6} & 7.56e-4 & 1.25e-5 & 2.76e-3 \\ \hline
			\multicolumn{1}{|r|}{low} & \multicolumn{1}{r|}{2.06e-4} & 4.54e-2 & 7.52e-4 & 1.65e-1 \\ \hline
		\end{tabular}%
	}
	\label{tab:data_price}
\end{table}

In order to derive a profitable data price, we can assume that the cost of performing settlement operations has to be equal to 2\% of the price for that data amount and that only 1 cube settlement is performed per day. We consider two examples: 1) an air quality monitoring application, with low data rate, running on a low-power wide-area network (LPWAN), such as LoRaWAN, that samples and transmit data every hour, resulting in 24 measurements per day; 2) a heart rate monitoring application (e.g., Fitbit), with sampling frequency of 1 second and 1 minute corresponding to a high and medium data rate, respectively.  
Table~\ref{tab:data_price} shows that data price ranges from 5.73e-8 Ether (\$ 1.26e-5) to 7.52e-4 Ether (\$ 0.165) depending on data transfer rate and on the data feed type selected. 
%More specifically, Table~\ref{tab:data_price} shows estimated data price for different use cases, such as a heart rate monitoring application (e.g., fitbit), with sampling frequency of 1 second (high data rate), 1 minute (medium data rate) and 1 hour (low data rate).   

\subsection{Discussion}
\label{sec:discussion}
The analysis above helped us to identify the feasibility of building a decentralized open and transparent accounting infrastructure, useful to create a fair data marketplace, where data price can evolve depending on data quality, demand and offer. To minimize the shared costs of running such an infrastructure, we observed how the Gas price can be tweaked at the cost of a lower transaction rate, leveraging the lack of real-time requirements for the settlement operations. Moreover, we demonstrated how the cubes architecture allow for scalability by reducing the settlement transaction frequency.
Nevertheless, we recognize that the estimated infrastructure costs are related to the current inflation in the Ether value, due to the large number of currently deployed general purpose Smart Contracts (raising up the Ether price of over 20 times in just one year). While we plan to perform similar analysis using different blockchain implementations like hyperledger (\url{hyperledger.org}), we anticipate that a decentralized trading infrastructure will require to fork a new dedicated Ethereum network, dedicated to contract settlement, with lower incentive fees for the miners. While keeping it open, we are confident that, due to the large amount of IoT data exchanges such a market will provide a viable business opportunity for miners even at lower transaction and incentive fees. 
\subsection{Related work}

%%%%
% sensors as service
%%%%
%\note{   \cite{Perera2014}  Perera, C., Zaslavsky, A., Christen, P., \& Georgakopoulos, D. (2014). Sensing as a service model for smart cities supported by Internet of Things. Transactions on Emerging Telecommunications Technologies, 25(1), 81–93. https://doi.org/10.1002/ett.2704}

%\note{ \cite{distefano2012sensing}  Distefano, S., Merlino, G., \& Puliafito, A. (2012). Sensing and actuation as a service: A new development for clouds. In Network Computing and Applications (NCA), 2012 11th IEEE International Symposium on (pp. 272–275). }

%\note{Banerjee, P., Friedrich, R., Bash, C., Goldsack, P., Huberman, B., Manley, J., … Veitch, A. (2011). Everything as a Service: Powering the New Information Economy. Computer, 44(3), 36–43. https://doi.org/10.1109/MC.2011.67}

The idea of considering data from IoT sensors as tradeable assets is closely related to that of \textit{Sensing as a Service} (SaaS) models, or even \textit{Sensing and Actuation as a service} (SAaaS) \cite{distefano2012sensing}, themselves derivatives of the more general ``Everything as a Service'' (XAAS) cloud-based model for data exchange \cite{5719575}. Perera et al. \cite{Perera2014}, for instance, outline a vision of a near future for Smart Cities, where data streams emanating from pervasive IoT sensors are accessible through services. The SaaS model consists of four conceptual layers: sensors and their owners, sensor publishers, service providers, and sensor data consumers. In this classification, our work is relevant to all of these agents, as it enables fair and metered data exchanges amongst sensors owners and publishers on one side, and sensor data consumers, on the other.
%
%%%%
%\note{related: data pricing}
%%%%
%\note{\cite{Li:2014:TPP:2691190.2691191} Li, C., Li, D. Y., Miklau, G., \& Suciu, D. (2014). A Theory of Pricing Private Data. ACM Trans. Database Syst., 39(4), 34:1--34:28. https://doi.org/10.1145/2691190.2691191}
%\note{\cite{7437020} Niyato, D., Hoang, D. T., Luong, N. C., Wang, P., Kim, D. I., \& Han, Z. (2016). Smart data pricing models for the internet of things: a bundling strategy approach. IEEE Network, 30(2), 18–25. https://doi.org/10.1109/MNET.2016.7437020}
%\note{related: trust management}
%\note{ \cite{Bao:2012:DTM:2378023.2378025} Bao, F., \& Chen, I.-R. (2012). Dynamic Trust Management for Internet of Things Applications. In Proceedings of the 2012 International Workshop on Self-aware Internet of Things (pp. 1–6). New York, NY, USA: ACM. https://doi.org/10.1145/2378023.2378025}
%\note{\cite{Yan2014a} Yan, Z., Zhang, P., \& Vasilakos, A. V. (2014). A survey on trust management for Internet of Things. Journal of Network and Computer Applications, 42, 120–134. https://doi.org/10.1016/j.jnca.2014.01.014}
%

Our traffic monitoring infrastructure assumes that suitable pricing models (covering at least the minimum transaction fees) that associate values to messages are in place.
However, it is agnostic and ``orthogonal'' to the specific pricing model, as long as the price of a complex bundle of data offering can be expressed in terms of unit cost associated to individual messages. 
Thus, in principle, any of the existing models for data pricing may be used in combination with traffic metering. Such models, recently proposed, range from theoretical frameworks for assigning prices to query answers as a function of their accuracy \cite{Li:2014:TPP:2691190.2691191}, to adaptations of \textit{Smart Data Pricing} \cite{Sen:2015:SDP:2847579.2756543} to the dynamic pricing of IoT data, such as personal data from wearable sensors \cite{7437020}.
A trust management model should also be established, i.e., to enable self-regulation of marketplace rules, as we briefly discussed.
While this is out of our scope, existing frameworks can be used on top of our infrastructure.
Yan et al \cite{Yan2014a} provide a starting point, by exploring the notion of trust across the IoT platform layers (physical sensing, network, and application layers), with the focus on a wide range of properties from security to goodness, strength, reliability, availability, ability of data. However, their survey largely overlooks issues of trust amongst participants in a data marketplace, i.e., in the context of data exchange transactions.
%
%\note{ Roman, D., \& Stefano, G. (2016). Towards a Reference Architecture for Trusted Data Marketplaces: The Credit Scoring Perspective. In 2016 2nd International Conference on Open and Big Data (OBD) (pp. 95–101). IEEE. https://doi.org/10.1109/OBD.2016.21}
More directly useful in our setting, as we progress in our work, is Roman and Gatti's study of trust in data marketplaces \cite{7573695}, based on \textit{credit scoring}, where a direct connection is made to the use of blockchain technology with data trading.
%
%%%%
%\note{related: marketplace for data}
%%%%

Two technical architectures for data marketplaces are directly relevant to our work.
%\note{ Schomm, F., Stahl, F., \& Vossen, G. (2013). Marketplaces for Data: An Initial Survey. SIGMOD Rec., 42(1), 15–26. https://doi.org/10.1145/2481528.2481532}
%\note{\cite{Cao:2016:MMR:2926746.2883611}  Cao, T.-D., Pham, T.-V., Vu, Q.-H., Truong, H.-L., Le, D.-H., \& Dustdar, S. (2016). MARSA: A Marketplace for Realtime Human Sensing Data. ACM Trans. Internet Technol., 16(3), 16:1--16:21. https://doi.org/10.1145/2883611}
Firstly, the MARSA platform \cite{Cao:2016:MMR:2926746.2883611}, designed specifically to deal with real-time data streams by interacting with existing IoT platforms.
The motivation for this work is very similar to ours, namely to provide a marketplace where owners have an incentive to trade their data, for either personal or community benefit.
The many technical requirements that emerge from the analysis of the data marketplace potential translate into a complex architecture, which includes data flow orchestration, participants registration, data contract management, and payment.  
%While these components do address complex marketplace requirements, our main challenge, namely to remove the need for a central trusted authority to manage the marketplace and ensure its fairness, remains unique to our work.
While these components do address complex marketplace requirements, the challenge to remove the need for a central trusted authority to manage the marketplace and ensure its fairness remains unique to our work.
Secondly, Misura, K., \& Zagar \cite{7765669} focus on a query-based mechanism, whereby devices register their data supply capabilities to a broker along with a number of properties, and consumers express interest in data streams by querying those properties. The broker then connects the supplier stream to the consumer, and monitors usage. 
This is relevant work, as this type of matching of consumer data requirements to suppliers capabilities is more sophisticated than simple topic subscription. Our work is complementary to this  and also contributes to remove the trust from the broker for monitoring usage.
In our future work, we plan to move away from a fine-grained data subscription and towards complex data contracts (bundles).
%\note{ Misura, K., \& Zagar, M. (2016). Data marketplace for Internet of Things. In 2016 International Conference on Smart Systems and Technologies (SST) (pp. 255–260). IEEE. https://doi.org/10.1109/SST.2016.7765669}
%\note{\cite{7146004} Blazquez, A., Tsiatsis, V., \& Vandikas, K. (2015). Performance Evaluation of OpenID Connect for an IoT Information Marketplace. In 2015 IEEE 81st Vehicular Technology Conference (VTC Spring) (pp. 1–6). IEEE. https://doi.org/10.1109/VTCSpring.2015.7146004}	
%\input{realisation}
%\vspace{-15pt}
\section{Conclusion and future work}
Our initial work encourages us to further develop the idea of a decentralized data marketplace, where benefits such as interoperability, transparency and fairness are achievable and cost affordable. However, we recognize that the cube settlement component is a very important but still only one building block of such an infrastructure, in which existing less-critical centralized and new decentralized elements will have to be combined. In the future, we plan to experiment and test the effectiveness of the reputation based reconciliation strategy and to develop the complete architecture for a trusted and transparent data marketplace. This will include data producers and VASs discovery service, contract creation and discovery platform, and the definition of an open governance model associated to it, promoting public and open creation, and review of settlement contracts (extending the github model (\url{github.com})).


\section*{Acknowledgment}
The authors would like to thank Ms. Lucie Burgess from the Digital Catapult Centre in London for her help in researching data marketplaces.
This work has been partially supported by the H2020 project SynchroniCity No 732240.

%\bibliographystyle{IEEEtran}
%\bibliography{IEEEabrv,iot-conf}

\small %\footnotesize



%\bibliographystyle{IEEEtran}

\begin{thebibliography}{10}
\providecommand{\url}[1]{#1}
\csname url@samestyle\endcsname
\providecommand{\newblock}{\relax}
\providecommand{\bibinfo}[2]{#2}
\providecommand{\BIBentrySTDinterwordspacing}{\spaceskip=0pt\relax}
\providecommand{\BIBentryALTinterwordstretchfactor}{4}
\providecommand{\BIBentryALTinterwordspacing}{\spaceskip=\fontdimen2\font plus
\BIBentryALTinterwordstretchfactor\fontdimen3\font minus
  \fontdimen4\font\relax}
\providecommand{\BIBforeignlanguage}[2]{{%
\expandafter\ifx\csname l@#1\endcsname\relax
\typeout{** WARNING: IEEEtran.bst: No hyphenation pattern has been}%
\typeout{** loaded for the language `#1'. Using the pattern for}%
\typeout{** the default language instead.}%
\else
\language=\csname l@#1\endcsname
\fi
#2}}
\providecommand{\BIBdecl}{\relax}
\BIBdecl

\bibitem{7004800}
C.~Perera, C.~H. Liu, and S.~Jayawardena, ``{The Emerging Internet of Things
  Marketplace From an Industrial Perspective: A Survey},'' \emph{IEEE
  Transactions on Emerging Topics in Computing}, vol.~3, no.~4, pp. 585--598,
  dec 2015.

\bibitem{7113786}
S.~M.~R. Islam, D.~Kwak, M.~H. Kabir, M.~Hossain, and K.~S. Kwak, ``{The
  Internet of Things for Health Care: A Comprehensive Survey},'' \emph{IEEE
  Access}, vol.~3, pp. 678--708, 2015.

\bibitem{Perera2014}
C.~Perera, A.~Zaslavsky, P.~Christen, and D.~Georgakopoulos, ``{Sensing as a
  service model for smart cities supported by Internet of Things},''
  \emph{Transactions on Emerging Telecommunications Technologies}, vol.~25,
  no.~1, pp. 81--93, jan 2014.

\bibitem{Stahl2016}
F.~Stahl, F.~Schomm, G.~Vossen, and L.~Vomfell, ``{A classification framework
  for data marketplaces},'' \emph{Vietnam Journal of Computer Science}, vol.~3,
  no.~3, pp. 137--143, aug 2016.

\bibitem{7765669}
K.~Misura and M.~Zagar, ``{Data marketplace for Internet of Things},'' in
  \emph{2016 International Conference on Smart Systems and Technologies
  (SST)}.\hskip 1em plus 0.5em minus 0.4em\relax IEEE, oct 2016, pp. 255--260.

\bibitem{Schomm2013}
F.~Schomm, F.~Stahl, and G.~Vossen, ``{Marketplaces for Data: An Initial
  Survey},'' \emph{SIGMOD Rec.}, vol.~42, no.~1, pp. 15--26, 2013.

\bibitem{Cao:2016:MMR:2926746.2883611}
T.-D. Cao, T.-V. Pham, Q.-H. Vu, H.-L. Truong, D.-H. Le, and S.~Dustdar,
  ``{MARSA: A Marketplace for Realtime Human Sensing Data},'' \emph{ACM Trans.
  Internet Technol.}, vol.~16, no.~3, pp. 16:1----16:21, may 2016.

\bibitem{Buterin2014}
\BIBentryALTinterwordspacing
V.~Buterin, ``{A next-generation smart contract and decentralized application
  platform},'' 2014. [Online]. Available:
  \url{http://buyxpr.com/build/pdfs/EthereumWhitePaper.pdf}
\BIBentrySTDinterwordspacing

\bibitem{distefano2012sensing}
S.~Distefano, G.~Merlino, and A.~Puliafito, ``{Sensing and actuation as a
  service: A new development for clouds},'' in \emph{Network Computing and
  Applications (NCA), 2012 11th IEEE International Symposium on}.\hskip 1em
  plus 0.5em minus 0.4em\relax IEEE, 2012, pp. 272--275.

\bibitem{5719575}
P.~Banerjee, R.~Friedrich, C.~Bash, P.~Goldsack, B.~Huberman, J.~Manley,
  C.~Patel, P.~Ranganathan, and A.~Veitch, ``{Everything as a Service: Powering
  the New Information Economy},'' \emph{Computer}, vol.~44, no.~3, pp. 36--43,
  mar 2011.

\bibitem{Li:2014:TPP:2691190.2691191}
C.~Li, D.~Y. Li, G.~Miklau, and D.~Suciu, ``{A Theory of Pricing Private
  Data},'' \emph{ACM Trans. Database Syst.}, vol.~39, no.~4, pp. 34:1----34:28,
  dec 2014.

\bibitem{Sen:2015:SDP:2847579.2756543}
S.~Sen, C.~Joe-Wong, S.~Ha, and M.~Chiang, ``{Smart Data Pricing: Using
  Economics to Manage Network Congestion},'' \emph{Commun. ACM}, vol.~58,
  no.~12, pp. 86--93, nov 2015.

\bibitem{7437020}
D.~Niyato, D.~T. Hoang, N.~C. Luong, P.~Wang, D.~I. Kim, and Z.~Han, ``{Smart
  data pricing models for the internet of things: a bundling strategy
  approach},'' \emph{IEEE Network}, vol.~30, no.~2, pp. 18--25, mar 2016.

\bibitem{Yan2014a}
Z.~Yan, P.~Zhang, and A.~V. Vasilakos, ``{A survey on trust management for
  Internet of Things},'' \emph{Journal of Network and Computer Applications},
  vol.~42, pp. 120--134, jun 2014.

\bibitem{7573695}
D.~Roman and G.~Stefano, ``{Towards a Reference Architecture for Trusted Data
  Marketplaces: The Credit Scoring Perspective},'' in \emph{2016 2nd
  International Conference on Open and Big Data (OBD)}.\hskip 1em plus 0.5em
  minus 0.4em\relax IEEE, aug 2016, pp. 95--101.

\bibitem{7553037}
D.~Niyato, X.~Lu, P.~Wang, D.~I. Kim, and Z.~Han, ``{Economics of Internet of
  Things: an information market approach},'' \emph{IEEE Wireless
  Communications}, vol.~23, no.~4, pp. 136--145, aug 2016.

\bibitem{nakamoto2008bitcoin}
S.~Nakamoto, ``Bitcoin: A peer-to-peer electronic cash system,'' 2008,
  \url{https://bitcoin.org/bitcoin.pdf}.

\bibitem{schaub2016trustless}
A.~Schaub, R.~Bazin, O.~Hasan, and L.~Brunie, ``A trustless privacy-preserving
  reputation system,'' in \emph{IFIP International Information Security and
  Privacy Conference}.\hskip 1em plus 0.5em minus 0.4em\relax Springer, 2016,
  pp. 398--411.

\bibitem{carboni2015feedback}
D.~Carboni, ``Feedback based reputation on top of the bitcoin blockchain,''
  \emph{arXiv preprint arXiv:1502.01504}, 2015.

\end{thebibliography}

\end{document}